\documentclass[aip,jcp,reprint,amsmath,amssymb]{revtex4-2}
\usepackage{graphicx}
\usepackage[suffix=]{epstopdf}
\usepackage{natmove}
\usepackage{amsmath,amssymb}

\renewcommand{\d}[1]{{\operatorname{d}\!{#1}}}

\begin{document}
\title{Axial dispersion in dilute solutions of linear and branched polymers in parallel-plate and expansion--contraction microchannels}

\author{C. Levi Petix}
\thanks{These authors contributed equally.}
\affiliation{Department of Chemical Engineering, Auburn University, Auburn, AL 36849, USA}

\author{Tzortzis Koulaxizis}
\thanks{These authors contributed equally.}
\affiliation{Department of Chemical and Biomolecular Engineering, University of Illinois, Urbana--Champaign, Illinois 61801, USA}

\author{Griffin D. Overton}
\affiliation{Department of Chemical Engineering, Auburn University, Auburn, AL 36849, USA}

\author{Antonia Statt}
\email{statt@illinois.edu}
\affiliation{Department of Chemical and Biomolecular Engineering, University of Illinois, Urbana--Champaign, Illinois 61801, USA}
\affiliation{Department of Materials Science and Engineering, The Grainger College of Engineering,  University of Illinois, Urbana--Champaign, Illinois 61801, USA}

\author{Michael P. Howard}
\email{mphoward@auburn.edu}
\affiliation{Department of Chemical Engineering, Auburn University, Auburn, AL 36849, USA}

\begin{abstract}
The axial dispersion of polymers in microchannels depends on an interplay between microchannel geometry, polymer architecture, and hydrodynamics. Here, we investigate the axial dispersion of linear, comb, and star polymers in parallel‑plate and sinusoidal expansion--contraction microchannels at dilute concentrations using multiparticle collision dynamics simulations. The polymers all contain the same number of monomers but differ in their architecture, and their concentration is fixed at either one value that is dilute for all polymers or the same value relative to the overlap concentration for each polymer. The dispersion coefficients measured at a nominal solvent volumetric flow rate are found to depend on both architecture and concentration. We show that the dispersion coefficients collapse as a function of the P\'{e}clet number after accounting for confinement effects on the polymer diffusion coefficient and polymer contributions to the flow field, and the dispersion coefficients in the parallel-plate microchannel can be reasonably predicted using a theory that accounts for inhomogeneous distribution of the polymers in the microchannel.
\end{abstract}

\maketitle

\section{Introduction}
The dispersion of polymers in flow is a fundamental transport process with practical implications for numerous applications, including mixing \cite{Locascio2004,Dentz_mixing}, filtration \cite{Nikoubashman_filter,Ollila_filtration}, and controlled delivery \cite{datta_dispersion_microfluidics}. Axial dispersion measurements can also be used to determine size \cite{Ibrahim2013} as well as to monitor polymerization reactions \cite{Cottet_2010}. The theoretical basis for understanding axial dispersion was established by Taylor \cite{Taylor,Taylor_1954} and Aris \cite{Aris}. Their analysis showed that the axial dispersion of point-like tracers in Hagen--Poiseuille (pipe) flow is enhanced compared to diffusion alone because of migration across streamlines, which causes the tracers to experience local velocities different from the mean velocity and spread. For polymers in microchannels, however, axial dispersion can be complicated by the geometry of the microchannel and the molecular structure of the polymer \cite{stone_microfluidics,squires_fluid_physics,ajdari_dispersion_microchannels}.

The geometry of the microchannel impacts the streamlines and local velocities that can be sampled. For example, flow in sinusoidal expansion--contraction microchannels can differ substantially from that in a constant cross-section microchannel with the same average size \cite{Kitanidis1997,newtonian_svolent_flow}, and as a result, the dispersion of point-like tracers in periodic expansion--contraction microchannels depends on both the length, the average width, and the oscillation amplitude of the microchannel \cite{Hoagland_1985, Bolster_solute_apertures}. Other studies have quantified how dispersion changes in patterned~\cite{Bhaumik2015251}, sinusoidal~\cite{Androver_laminar}, arbitrarily asymmetric~\cite{Chang2022TaylorDI,teodoro2025taylor}, and corrugated microchannels~\cite{Alexandre_corrugated} because of variations in geometry. Axial dispersion in microchannels with varying cross sections can be analyzed using extensions of the Taylor--Aris theory \cite{Brenner_1980, Rosecrans_1997}; however, these analyses have typically focused on point-like tracers that fully explore the microchannel and do not modify the underlying flow.

The molecular nature of polymers presents several challenges compared to point-like tracers. First, polymers have finite size and so their centers of mass are effectively excluded from exploring streamlines in certain regions of the microchannel. Such exclusion has been shown to significantly affect dispersion of spherical colloidal particles \cite{Brenner_brownian,dispersion_brownian}. Second, polymers can undergo flow-induced deformation and redistribute in the microchannel, leading to changes in how they sample streamlines that depend on the flow rate \cite{Li_star_branched_dpd, Yang2020}. For example, linear polymers can migrate in flow and adopt distributions different than at equilibrium \cite{migration_depablo}. Third, polymers contribute to the solution viscosity and can impart non-Newtonian behavior even at modest concentrations, modifying the flow field compared to a Newtonian solvent \cite{Vartuli_capillary}.  As a result of these considerations, polymer dispersion may depend not only on the flow rate and the microchannel geometry as for point-like solutes, but also on the molecular structure of the polymer, including size and functionality\cite{Nikoubashman_branched, Ripoll_star, Ripoll_shear, Nikoubashman_filter}. For example, star polymers\cite{Nikoubashman_chain_star_mixtures,Liu2020RoleOF} and ring polymers\cite{weiss2019spatial} migrate more toward the center of a parallel-plate microchannel than linear polymers of comparable size.

Theoretical analysis of the dispersion of even an idealized elastic-dumbbell polymer in simple flow geometries is surprisingly complex \cite{park_dispersion}, making numerical simulations a powerful strategy for understanding the dynamics of polymers in microchannels \cite{Li_star_branched_dpd,weiss2019hydrodynamics,Yang2020,migration_depablo,Nikoubashman_chain_star_mixtures,Liu2020RoleOF,weiss2019spatial}. Here, we have employed mesoscale particle-based simulations to investigate the axial dispersion of solutions of linear and branched polymers in a parallel-plate and a sinusoidal expansion--contraction microchannel. The polymers all consisted of the same number of monomers to facilitate comparison across three different polymer architectures: linear, comb, and star. We considered dilute polymer solutions with either the same concentration for all polymers or the same concentration relative to their respective overlap concentrations. Despite differences in polymer size, polymer deformability, and apparent solution viscosity, we found that the axial dispersion coefficient could be collapsed as a function of the P\'{e}clet number defined using the polymer diffusion coefficient and the average velocity of the polymer solution for the conditions studied. These findings are useful for understanding and analyzing the dynamics of polymers in microchannels.

\section{Model and Methods}
\label{sec:model}
All quantities will be reported in an arbitrary consistent system of units where $m$ is the unit of mass, $\ell$ is the unit of length, and $\varepsilon$ is the unit of energy. The unit of time is $\tau = \sqrt{m\ell^2/\varepsilon}$, and the unit of temperature is $\varepsilon/k_{\rm B}$ where $k_{\rm B}$ is the Boltzmann constant. Simulations were performed using HOOMD-blue \cite{hoomd} (version 5.4.0) extended with azplugins \cite{azplugins} (version 1.2.0) using a timestep of $0.005\,\tau$ and were repeated three times using different initial configurations. We report the average of quantities measured from these independent simulations with uncertainty estimated as one standard error between the measurements.

\subsection{Polymer solutions}
Bead--spring polymers were modeled using the Kremer--Grest approach \cite{Kg_original}. All monomers had diameter $d=1\,\ell$ and interacted through the purely repulsive Weeks--Chandler--Andersen (WCA) pair potential \cite{Weeks_wca_1971},
\begin{equation}
U_{\rm WCA}(r)=
\begin{cases}
4\varepsilon\left[\left(\dfrac{d}{r}\right)^{12}-\left(\dfrac{d}{r}\right)^6 + \dfrac{1}{4}\right], & r\le 2^{1/6}d \\
0, & {\rm otherwise}
\end{cases},
\label{eq:wca}
\end{equation}
where $r$ is the distance between bead centers. Bonded monomers additionally interacted through a finitely extensible nonlinear elastic (FENE) potential,
\begin{equation}
U_{\rm FENE}(r)=-\frac{1}{2}kr_0^2\ln\left[1-\left(\frac{r}{r_0}\right)^2\right],
\label{eq:fene}
\end{equation}
with spring constant $k = 30\,\varepsilon/\ell^2$ and maximum bond length $r_0 = 1.5\,\ell$.

We studied three polymer architectures: linear, comb, and star. All polymers had 97 monomers that were connected in different topologies. The linear polymer consisted of 97 sequentially bonded monomers. The comb polymer consisted of a 49-monomer linear backbone with $f$ linear arms of $48/f$ monomers attached at regular intervals (Fig.~S1). The star polymer consisted of a central monomer bonded to $f$ linear arms of $96/f$ monomers. We considered comb polymers with $f=4$, 6, 8, 16, and 24 and star polymers with $f=4$, 6, 8, and 12.

The solvent was modeled using multiparticle collision dynamics (MPCD) \cite{gompper_2008_mpcd_review,howard_modeling_2019, Kapral_review}. The solvent consisted of point particles with mass $1\,m$ that did not have pairwise interactions; instead, their dynamics propagated in alternating streaming and collision steps. Specifically, the solvent particles streamed according to Newton's equations of motion for $0.1\,\tau$ in a single step, then they were sorted into cubic cells of edge length $1\,\ell$ to perform the collision. Particles exchanged momentum with other particles in the same cell using the stochastic rotation dynamics rule without angular momentum conservation \cite{malevanets_1999_mesoscopic}. We used a fixed rotation angle of $130^{\circ}$, a rotation axis drawn randomly from the unit sphere, and a cell-level Maxwell--Boltzmann thermostat \cite{huang_thermostat_mpcd} with temperature $T=1\,\varepsilon/k_{\rm B}$. The collision cells were shifted at each collision step by a random vector whose components were uniformly distributed in $[-\ell/2,\ell/2]$ to ensure Galilean invariance \cite{ihle_2001_galilean_srd}. The solvent number density was $10\,\ell^{-3}$, giving a theoretically expected dynamic viscosity of $8.70\,\varepsilon\tau/\ell^3$ \cite{Ripoll_dynamic}. The monomers were coupled to the solvent by participating in the collision step \cite{malevanets_2000,mussawisade_solvent}. Between collisions, the monomers were propagated according to Newton's equations of motion using velocity Verlet integration \cite{Allen}. The monomer mass was set to $10\,m$ to match the average mass of solvent in a collision cell \cite{Ripoll_dynamic,mussawisade_solvent}.

We simulated dilute polymer solutions having concentration $c = N/V$, where $N$ is the number of polymers and $V$ is the volume. The polymers were expected to have different sizes due to their different architectures, so we considered two schemes for choosing the concentration: one in which $c$ was fixed at a value that was considered dilute for all polymers and one in which $c$ was fixed relative to the overlap concentration $c^* = 3/(4 \pi R_{\rm g}^3)$ for each polymer, where $R_{\rm g}$ is the root mean squared radius of gyration of the polymer at infinite dilution. To determine $c^*$, we measured $R_{\rm g}$ using implicit-solvent Langevin dynamics simulations with friction coefficient $0.1\,m/\tau$ \cite{Allen}, which produce the same equilibrium polymer structures as in the MPCD solvent but with less compute time. The monomer mass was $1\,m$ for these simulations. We randomly placed ten polymers without overlap in a cubic simulation box with side length $200\,\ell$ and periodic boundary conditions. After an equilibration period of $10^4\,\tau$, polymer configurations were sampled every $10\,\tau$ during a $10^5\,\tau$ production period, and $R_{\rm g}$ was calculated by averaging over all polymers in all sampled configurations. We then calculated the concentration $c$ giving $c/c^* = 0.5$, which is considered dilute, for each polymer and used these concentrations for the simulations at fixed $c/c^*$. The linear polymers had the largest $R_{\rm g}$, so this concentration was also dilute for all polymers and used for the simulations with fixed $c$.

We then characterized the bulk polymer self-diffusion coefficients $D$ for all architectures at these concentrations. For the polymer solutions with fixed $c$, we used a cubic simulation box with side length $100\,\ell$ and periodic boundary conditions, giving 305 polymers. For the solutions of comb and star polymers with fixed $c/c^*$, the box size was adjusted to keep the number of polymers approximately the same as for the fixed $c$ simulations (Table S1). After an equilibration period of $10^5\,\tau$, polymer configurations were recorded every $10^3\,\tau$ during a $2\times10^6\,\tau$ production period. The mean squared displacement $\langle \Delta r^2\rangle$ of the center of mass of a polymer was calculated for times $t \le 2 \times 10^5 \,\tau$ using all configurations as time origins and averaging over all polymers. The self-diffusion coefficient $D$ was then extracted using the Einstein relation,
\begin{equation}
D= \lim_{t \to \infty} \frac{1}{6}\frac{{\rm d}\langle\Delta r^2\rangle}{{\rm d}t}.
\end{equation}
The time derivative was calculated numerically, and the limit was taken by averaging the time derivative over $t \ge 10^5 \,\tau$, where it was observed to be roughly constant.

\subsection{Microchannels}
We studied polymer dispersion in both a parallel-plate microchannel and an expansion--contraction microchannel using the same geometry studied by Kitanidis and Dykaar \cite{Kitanidis1997} and in our prior work \cite{newtonian_svolent_flow}. The flow direction was $x$, confining walls were placed in the $y$ direction, and the $z$ direction was unconfined. The half-width $H$ between the channel walls, which were centered around the origin, was
\begin{equation}
H(x)=\frac{W}{2}+A\cos\left(\frac{2\pi x}{L}\right),
\label{eq:walls}
\end{equation}
where $L$ is the length of the microchannel, $W$ is the average distance between the walls, and $A$ is the wall amplitude. The microchannel had depth $D$ in the $z$ direction, and periodic boundary conditions were used in the $x$ and $z$ directions. The parallel-plate microchannel corresponds to setting $A=0$. Both the solvent and the monomers were reflected from the walls using bounce-back rules for no-slip boundary conditions based on their positions and velocities \cite{Lamura2001b, Whitmer_2010}, but the walls were shifted inward for the monomers by $\ell/2$ to account for their finite size. Virtual solvent particles were also added to the walls and participated in the collision to help enforce the no-slip boundary conditions \cite{Lamura2001b, Whitmer_2010}.

We initialized polymers in a microchannel with $W=60\,\ell$, $L=150\,\ell$, and depth $150\,\ell$, then the simulation box was compressed in the $z$ direction to depth $D=30\,\ell$ at a constant rate over a $10^5 \, \tau$ period. The number of polymers was chosen to achieve the desired concentration $c$ in the final microchannel geometry (Table S2). The system was then equilibrated for $10^5\,\tau$, after which 3 configurations were collected at intervals of $10^4\,\tau$. During these equilibration simulations, no solvent was included, the temperature was controlled using an isokinetic thermostat applied every $10^2\,\tau$, and the monomer mass was $1\,m$. These configurations were used as initial conditions for subsequent simulations.

Solvent was then added to the microchannel, and a constant body force $F$ was applied in the $x$ direction to both the solvent particles and the monomers to generate flow. The force was selected to achieve a nominal volumetric flow rate $Q$ of the pure solvent, which we determined for the parallel-plate microchannel using the well-known result for pressure-driven flow and for the expansion--contraction microchannel using the tenth-order series solution we previously derived (Table S3) \cite{newtonian_svolent_flow}. Flow was simulated for $2\times10^5\,\tau$ to reach steady state followed by a production period of $2\times10^6\,\tau$. During the production period, polymer configurations were recorded every $10^3\,\tau$, and the two-dimensional time- and mass-averaged velocity field $u_x(x,y)$ was computed every $10^2\,\tau$ by sorting particles into square bins of edge length $0.5\,\ell$.

The axial dispersion coefficient $K$ was obtained from the mean squared displacement of the center of mass of a polymer in the $x$ direction $\langle \Delta x^2\rangle$ relative to the mean displacement in the same direction $\langle \Delta x\rangle$. We determined these averages using all configurations as time origins and all polymers. We then calculated the second central moment $M_2 = \langle\Delta x^2\rangle - \langle \Delta x\rangle^2$ and numerically evaluated its time derivative $M_2'$. The axial dispersion coefficient was defined as
\begin{equation}
K = \lim_{t\to\infty} \frac{1}{2} M_2'.
\end{equation}
The long-time limit of $M_2'$ could not be directly observed in many of our simulations within the feasible computational time because reaching this limit requires the polymers to fully explore the flow field. However, for the parallel-plate microchannel, we found that $M_2'$ relaxed approximately as a decaying exponential,
\begin{equation}
M_2' = 2K + c_1 e^{-t/\tau_1},
\end{equation}
where $\tau_1$ is a relaxation time and $c_1$ is a fitting constant. For the expansion--contraction microchannel, $M_2'$ exhibited an additional damped oscillatory component, which we therefore fit to
\begin{equation}
M_2' = 2K + c_1 e^{-t/\tau_1} + c_2 e^{-t/\tau_2}\sin\left(\frac{2\pi t}{\tau_3}+c_3\right),
\end{equation}
where $c_2$ is the oscillation amplitude, $\tau_2$ is the damping time, $\tau_3$ is the oscillation period, and $c_3$ is a phase shift. The exceptions to this procedure for the expansion--contraction microchannel were the linear polymer with $Q = 25\,\ell^3/\tau$ as well as the comb and star polymers with $c$ held fixed and $Q = 20\,\ell^3/\tau$ and $25\,\ell^3/\tau$, which were better fit by the single decaying exponential. We performed these fits using SciPy \cite{Virtanen2020} (version 1.16.3) and recorded $K$ as the long-time dispersion coefficient (Fig.~S2).

We validated our simulation methods using solutions of monomers in the parallel-plate microchannel because theoretical predictions are available for both point-like and spherical solutes in that geometry \cite{Taylor,Aris,dispersion_brownian}. Specifically, we considered solutions of monomers obtained by removing the bonds from the linear polymers and from the 8-arm star polymers with fixed $c/c^*$, which had the smallest and largest number of monomers, respectively. The measured dispersion coefficients (Fig.~S3) followed the well-known Taylor--Aris prediction reasonably well when using the monomer self-diffusion coefficients measured in the parallel-plate geometry (Fig.~S4), but the simulated values were somewhat smaller. Accordingly, we also applied an extended theory  [see eq.~\eqref{eq:tayloraris} in Sec.~\ref{sec:results:parallel}] developed for confined Brownian colloids in parallel-plate microchannels \cite{dispersion_brownian} that accounts for confinement effects on particle distribution (Fig.~S5) and flow velocity (Fig.~S6), finding near-quantitative agreement when using simulated values as inputs. This comparison confirms that the MPCD simulations captured the physics of dispersion for simple solutes reasonably well, giving us confidence to apply them to the more complicated case of polymer dispersion.

\section{Results and Discussion}
\label{sec:results}
\subsection{Bulk properties}
\begin{figure}
    \centering
    \includegraphics{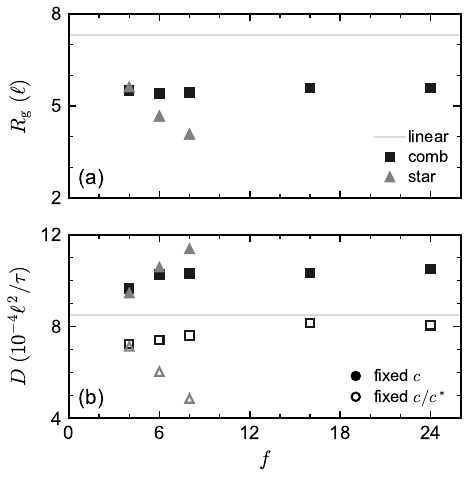}
    \caption{(a) Root mean squared radius of gyration $R_{\rm g}$ of polymers at infinite dilution. (b) Self-diffusion coefficient $D$ of polymers in dilute solutions with either fixed $c$ or fixed $c/c^*$. Results for the comb and star polymers are shown as functions of number of arms $f$. The uncertainty is approximately the size of a marker.}
    \label{fig:bulk}
\end{figure}

We first characterized the equilibrium root mean squared radius of gyration $R_{\rm g}$ of the polymers at infinite dilution and the self-diffusion coefficient $D$ of the polymers in bulk dilute solution (Fig.~\ref{fig:bulk}) to help interpret the dispersion coefficients (Sec.~\ref{sec:results:dispersion}). We emphasize that all polymers had the same number of monomers and differed only in their architecture. The linear polymer had the largest $R_{\rm g}$ [Fig.~\ref{fig:bulk}(a)], which was expected because all 97 monomers were arranged along a single backbone. The comb polymers were expected and found to be smaller than the linear polymer, but interestingly, $R_{\rm g}$ depended only weakly on the number of arms $f$. The number of monomers in the backbone was constant, so redistributing the monomers between a small number of long arms and a large number of short arms did not substantially alter the overall size of the combs considered here. This is consistent with prior work \cite{Mai_combs} showing that $R_{\rm g}$ scales primarily with the stretching factor, the product of the grafting density and degree of polymerization of the arms: since this quantity remained constant for all the combs considered here, they had similar overall sizes. In contrast, the star polymers became progressively more compact as $f$ increased because the individual arms became shorter and the overall polymer shape became more spherical.

The self-diffusion coefficient $D$ of the polymers in dilute solution also varied with $f$, but its behavior was different depending on whether $c$ or $c/c^*$ was fixed [Fig. \ref{fig:bulk}(b)]. At fixed $c$, $D$ increased for the star polymers as $f$ increased and $R_{\rm g}$ decreased, which is qualitatively consistent with the generalized Stokes--Einstein relationship \cite{mason:physrevlett:1995} for an object diffusing through a medium with the same zero-shear viscosity. (We note that there may also be small differences in viscosity at fixed $c$ due to variations in the intrinsic viscosity with polymer architecture \cite{Schaefgen_viscosities,Burchard1999SolutionPO,zimm_branches}.) In contrast, at fixed $c/c^*$, $D$ decreased significantly for the star polymers as $f$ increased. This qualitative difference is due to an increase in the solution viscosity because $c^*$ (and hence $c$) increased for the star polymers as $f$ increased and $R_{\rm g}$ decreased. Last, $D$ initially increased for the comb polymers as $f$ increased, regardless of whether $c$ or $c/c^*$ was fixed, but then plateaued. Based on the behavior of the star polymers, we suspect this behavior is caused by $R_{\rm g}$ remaining nearly constant and a decreasing influence of the arms on the solution viscosity as $f$ increased and the arms became very short.

Hence, the star polymers were more sensitive to the number of arms $f$ than the comb polymers, both structurally and dynamically. Based on their bulk properties, we decided to focus the dispersion simulations on a representative subset of polymers: the linear polymer, the 4-arm and 8-arm comb polymers, and the 4-arm and 8-arm star polymers. These choices span the observed polymer sizes and diffusion coefficients well, while avoiding architectures that differed only slightly in their bulk properties (e.g., 16-arm and 24-arm comb polymers).

\subsection{Dispersion coefficients}
\label{sec:results:dispersion}
\begin{figure*}
    \centering
    \includegraphics{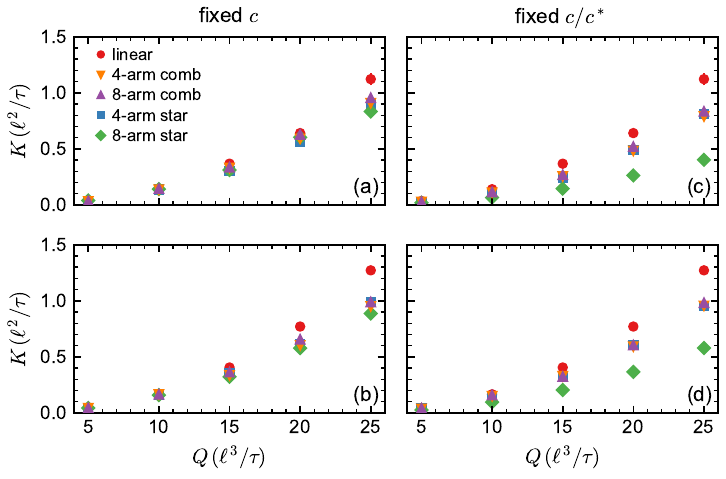}
    \caption{Axial dispersion coefficient $K$ for polymers at varying nominal flow rates $Q$ in (a, c) the parallel-plate microchannel and (b, d) the expansion--contraction microchannel with $A/W=0.2$. (a) and (b) are for fixed concentration $c$, while (c) and (d) are for fixed concentration relative to overlap, $c/c^*$. The uncertainty is approximately the size of a marker.}
    \label{fig:raw_dispersion}
\end{figure*}

Next, we considered the axial dispersion coefficient $K$ for the polymers in the parallel-plate and expansion--contraction microchannels at either fixed $c$ or fixed $c/c^*$ (Fig.~\ref{fig:raw_dispersion}). In all cases, $K$ increased with the nominal volumetric flow rate $Q$. This trend was expected because stronger flow increases the magnitude of velocities sampled by the polymers; indeed, $K \sim Q^2$ in the classic Taylor--Aris picture of dispersion \cite{Taylor, Aris}. At a given $Q$, $K$ depended systematically on the polymer architecture, with the dispersion coefficients of both the comb and star polymers being consistently less than that of the linear polymer. Further, $K$ was nearly the same for the two comb polymers, which had similar sizes and diffusion coefficients in bulk solution, whereas it differed between the two star polymers, which had different sizes and diffusion coefficients in bulk solution. The dispersion coefficients were also usually somewhat larger in the expansion--contraction microchannel than in the parallel-plate microchannel for a given polymer at the same $Q$. These differences were more pronounced when $c/c^*$ was fixed rather than $c$, particularly for the star polymers.

\begin{figure}
    \centering
    \includegraphics{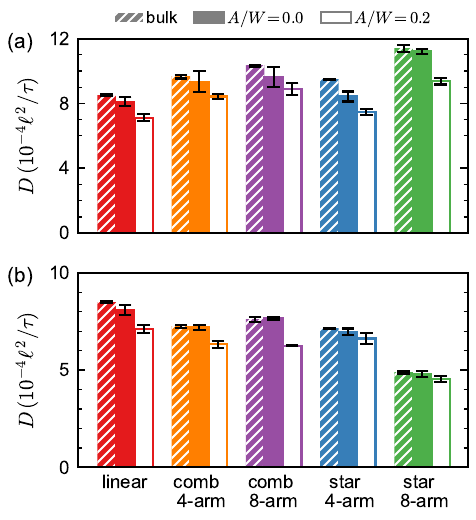}
    \caption{Self-diffusion coefficient $D$ for polymers in bulk solution, the parallel-plate microchannel ($A/W=0.0$), and the expansion--contraction microchannel ($A/W=0.2$). Results are shown for (a) fixed concentration $c$ and (b) fixed concentration relative to overlap, $c/c^*$.}
    \label{fig:diffusion}
\end{figure}

\begin{figure*}[!t]
    \centering
    \includegraphics{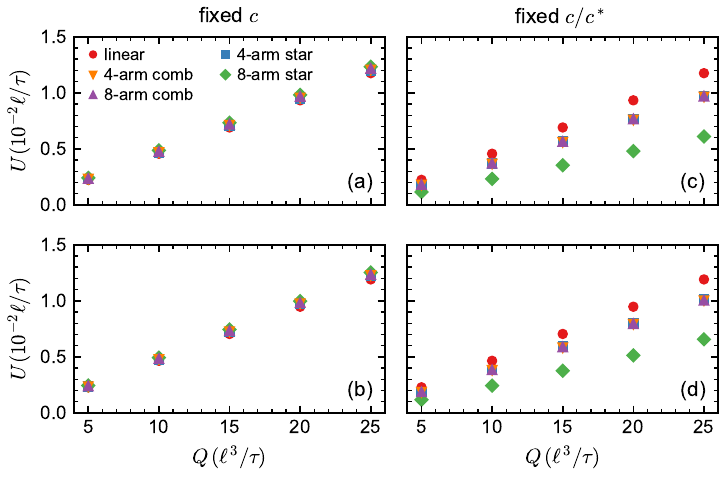}
    \caption{Average velocity $U$ of polymer solutions at varying nominal flow rates $Q$ in (a, c) the parallel-plate microchannel and (b, d) the expansion--contraction microchannel with $A/W=0.2$. (a) and (b) are for fixed concentration $c$, while (c) and (d) are for fixed concentration relative to overlap, $c/c^*$. The uncertainty is approximately the size of a marker.}
    \label{fig:solvent_velocities}
\end{figure*}

In interpreting Fig.~\ref{fig:raw_dispersion}, we emphasize some caution must be used in quantitatively comparing our measured values of $K$ at a given $Q$. The dispersion coefficient is expected to depend on the self-diffusion coefficient of the polymers, which varies with both architecture and concentration in bulk solution (Fig.~\ref{fig:bulk}), and may further depend on the microchannel geometry \cite{wang2023behaviors,chen_dynamics_dna,Doi_Harden}. Additionally, $Q$ is the nominal volumetric flow rate of the pure solvent, which we used as a controllable input parameter in our simulations, and not the actual volumetric flow rate of the polymer solution. The actual volumetric flow rate should depend on the solution viscosity and rheology (and so the polymer architecture), and differences in flow rate can lead to differences in $K$. 

We therefore first examined the polymer self-diffusion coefficient under confinement (Fig.~\ref{fig:diffusion}). Simulations of the self-diffusion coefficient in the microchannel were performed using the same protocol as for the dispersion coefficients but without a body force, and the diffusion coefficient was extracted by fitting a constant to $M_2'$. The diffusion coefficients in the parallel-plate microchannels were comparable to, but sometimes smaller than, those in bulk solution. Walls are well-known to reduce the diffusivity of spherical particles \cite{Brenner_1962,eral_anisotropic,Lin_isolated_sphere, Dettmer_diffusion}, and they also affect the self-diffusion coefficients of polymers \cite{wang2023behaviors,chen_dynamics_dna,Doi_Harden}. More strikingly, $D$ was consistently smaller in the expansion--contraction microchannel than in the parallel-plate microchannel, which is due to the enhanced confinement at the contraction point \cite{wang2023behaviors,diffusion_dna_strychalski}. The extent of this reduction depended on both the polymer architecture and the concentration, but it ranged from 4.9\% for the 4-arm star polymer to 18.5\% for the 8-arm comb polymer, both at fixed $c/c^*$. These results are important for interpreting the dispersion coefficients because they show that no single diffusion coefficient can be used for a given polymer, and instead one that accounts for geometry- and concentration-dependent changes is needed.

We then also examined the average velocity $U$ of the polymer solutions, which we calculated as a spatial average of the measured $u_x(x,y)$ using numerical integration (Fig.~\ref{fig:solvent_velocities}). The average velocity increased with $Q$ in all cases, as expected; however, it was not the same for all polymers at a fixed nominal $Q$, as would be the case for tracers. These differences were modest when $c$ was fixed, with the linear polymers having the largest $U$ and the 8-arm star polymers having the smallest $U$, but they were more pronounced when $c/c^*$ was fixed. These differences in $U$ with both polymer architecture and concentration must be due to differences in the solution rheology, as also implied by the measurements of $D$ (Fig.~\ref{fig:bulk}).

\begin{figure}
    \centering
    \includegraphics{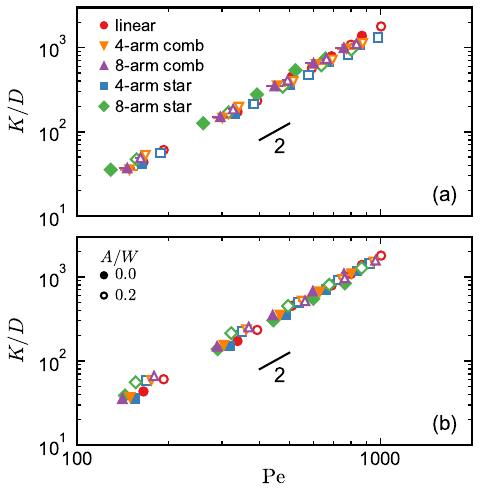}
    \caption{Dimensionless axial dispersion coefficient $K/D$ as a function of P\'{e}clet number Pe defined using the self-diffusion coefficient and solution velocity measured for each polymer and microchannel at (a) fixed concentration $c$ and (b) fixed concentration relative to overlap, $c/c^*$. The uncertainty is approximately the size of a marker.}
    \label{fig:collapse}
\end{figure}

Motivated by these observations and prior analysis of dispersion for point-like tracer particles \cite{Taylor,Aris,Frankel_Brenner_1989}, we attempted to collapse the dispersion coefficient as a function of the P\'{e}clet number ${\rm Pe} = U W/D$ using the self-diffusion coefficient $D$ and solution velocity $U$  measured for each polymer solution in each microchannel. The dimensionless dispersion coefficients $K/D$ seemed to collapse for the different polymers and microchannels (Fig.~\ref{fig:collapse}). The data for the different concentrations are presented on separate panels for clarity, but they essentially overlap. We also found that $K/D \sim {\rm Pe}^2$ as expected.

This collapse was somewhat surprising to us for two reasons. First, some of the polymers have different sizes, and for spherical tracer particles, differences in size lead to differences in the scaling prefactor of $K/D$ with ${\rm Pe}$ at large Pe \cite{dispersion_brownian,James_finite,Silebi_submicron}. Second, prior theoretical analysis has shown that point-like tracers are expected to have larger dispersion coefficients in our expansion--contraction microchannel than in the parallel-plate microchannel \cite{Bolster_solute_apertures}. Indeed, we ran simulations of monomer solutions in the expansion--contraction microchannel and confirmed that their dispersion coefficients did not collapse with those in the parallel-plate microchannel; they were typically about 46\% larger (Fig.~S7).

\subsection{Parallel-plate microchannel}
\label{sec:results:parallel}
To better understand the measured dispersion coefficients, we last investigated the behavior of the polymers in the parallel-plate microchannel. This geometry allowed us to extract spatially dependent quantities more easily and with more reliable statistics than the expansion--contraction microchannel. Additionally, a general theoretical analysis developed for particles using the Taylor--Aris approach can be applied for this geometry \cite{dispersion_brownian}. Specifically, the dispersion coefficient of an object isotropically diffusing in the velocity field $u_x(y)$ and having probability density $p(y)$ to be found at position $y$ is
\begin{equation}
\frac{K}{D} = 1 + \frac{1}{D^2} \int_{-W/2}^{W/2} \frac{1}{p(y)}\left(\int_{-W/2}^y [u_x(y') - \bar{u}_x] p(y') \d{y}' \right)^2 \d{y},
\label{eq:tayloraris}
\end{equation}
where
\begin{equation}
\bar{u}_x = \int_{-W/2}^{W/2} u_x(y) p(y) \d{y}
\end{equation}
is the average velocity of the object. Note that making lengths dimensionless using $W$ and velocities dimensionless using $U$ naturally introduces ${\rm Pe}$ into eq.~\eqref{eq:tayloraris}. We hence inspected both the velocity of the polymer solution and the distribution of polymers in the parallel-plate microchannel to identify similarities and differences for the different polymers and concentrations.

\begin{figure}
    \centering
    \includegraphics{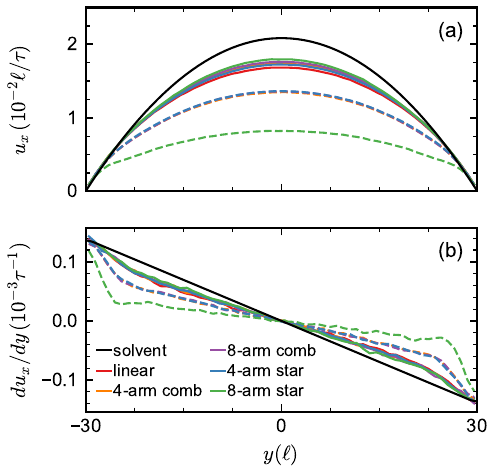}
    \caption{(a) Axial velocity profile $u_x(y)$ and (b) gradient of the axial velocity profile, $\mathrm{d}u_x/\mathrm{d}y$ in the parallel-plate microchannel for solutions of the linear polymer, 4-arm and 8-arm comb polymers, and 4-arm and 8-arm star polymers at nominal volumetric flow rate $Q=25\,\ell^3/\tau$. Results for fixed $c$ are shown with solid lines and results for fixed $c/c^*$ are shown with dashed lines. The expected behavior for the pure solvent is shown as a solid black line for reference.}
    \label{fig:velocity}
\end{figure}

We calculated $u_x(y)$ from the measured $u_x(x,y)$ at the highest flow rate by numerically integrating with respect to $x$ (Fig.~\ref{fig:velocity}), and several features were immediately apparent. First, the velocities of the polymer solutions were all less than the velocity of the pure solvent. Second, the velocities for the comb and star polymers with fixed $c/c^*$ were significantly smaller than with fixed $c$. Both of these features were expected based on Fig.~\ref{fig:solvent_velocities} and the concentration-dependent contribution of the polymers to the viscosity. Newly, the spatially resolved $u_x(y)$ for the polymer solutions also seemed to have a shape different from the parabolic form for the pure solvent. We confirmed that these deviations were not simply due to wall slip by also evaluating the gradient $\d{u_x}/\d{y}$, finding that it did not have the linear form for the pure solvent [Fig.~\ref{fig:velocity}(b)]. Qualitatively similar behavior was also found at lower flow rates in the parallel-plate microchannel (Fig.~S8) and at the expansion and contraction points of the expansion--contraction microchannel (Fig.~S9).

\begin{figure*}
    \centering
    \includegraphics{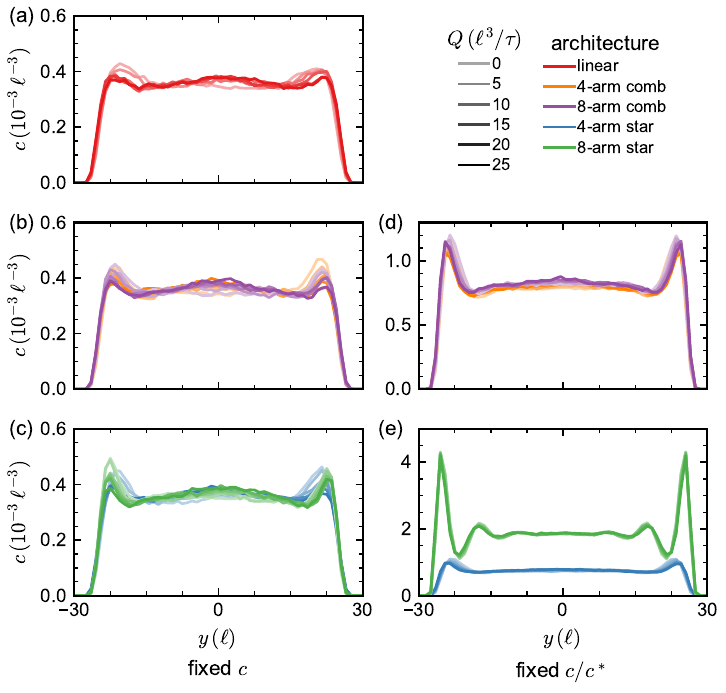}
    \caption{Concentration profile $c(y)$ of the polymer center of mass across the parallel-plate microchannel at fixed concentration $c$ for (a) the linear polymer, (b) the 4-arm and 8-arm comb polymers, and (c) the 4-arm and 8-arm star polymers and at fixed $c/c^*$ for (a) the linear polymer, (d) the 4-arm and 8-arm comb polymers, and (e) the 4-arm and 8-arm star polymers. In each panel, nominal volumetric flow rate $Q$ is denoted by increasing opacity.}
    \label{fig:polymer_distribution}
\end{figure*}

There may be two reasons for $u_x(y)$ becoming nonparabolic for the polymer solutions in the parallel-plate microchannel. First, polymer solutions can be non-Newtonian (e.g., shear thinning), leading to nonparabolic flow fields because of variations in the shear stress with varying shear rate across the microchannel. Further, because both the solvent and polymers experience the body force, the force density driving the flow has a spatial dependence if the polymers are inhomogeneously distributed across the microchannel. An inhomogeneous force density produces a nonparabolic velocity profile even for a Newtonian fluid. Both these reasons highlight the impact of the polymers on the flow field. Many theoretical treatments of dispersion assume that the solutes behave as tracers that do not alter the flow field, but polymers, even at modestly dilute concentrations, may alter it in ways that are architecture- and concentration-dependent.

To understand whether the polymers were inhomogeneously distributed, we then calculated the concentration $c(y)$ of the polymer center of mass across the parallel-plate microchannel and spatially averaged with respect to $x$ and $z$ (Fig.~\ref{fig:polymer_distribution}). Without flow ($Q=0\,\ell^3/\tau$), all polymers had a roughly uniform distribution for fixed $c$ [Fig.~\ref{fig:polymer_distribution}(a)--(c)], except for small peaks and a soft exclusion of size comparable to $R_{\rm g}$ near the walls. Without flow at fixed $c/c^*$, however, both the comb polymers and the 8-arm star polymer had more pronounced peaks near the walls [Fig.~\ref{fig:polymer_distribution}(d)--(e)]. The 8-arm star even had a layered distribution [Fig.~\ref{fig:polymer_distribution}(e)] that is reminiscent of confined hard spheres and so might be expected based on the sphere-like structure of this highly functionalized polymer. The distributions showed only a modest flow rate dependence, with increasing flow rate tending to decrease the peaks near the wall and form a new peak near the center of the microchannel. Such flow-induced migration has been noted previously \cite{Nikoubashman_chain_star_mixtures, weiss2019spatial,weiss2019hydrodynamics,Chelakkot_2011}. In addition to their indirect impact on $u_x(y)$, these inhomogeneous polymer distributions directly impact dispersion because they determine which parts of the flow field each polymer samples. A polymer that is broadly distributed across the channel usually samples a wider range of velocities and hence exhibits a larger dispersion than a polymer that is more narrowly distributed.

\begin{figure}
    \centering
    \includegraphics{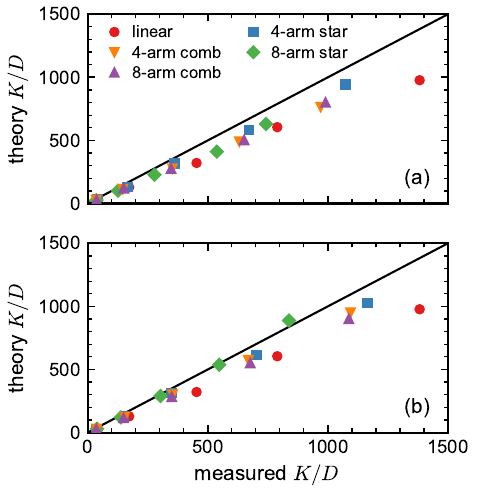}
    \caption{Comparison of the dimensionless dispersion coefficient $K/D$ in the parallel-plate microchannel predicted by eq.~\eqref{eq:tayloraris} using measured inputs to that measured directly in the simulations for (a) fixed concentration $c$ and (b) fixed concentration relative to overlap, $c/c^*$. The solid lines indicate equality of the measured and predicted values.}
    \label{fig:parity}

\end{figure}

To assess the direct impact of the polymer distribution on the dispersion, we applied eq.~\eqref{eq:tayloraris} using the measured $D$ and $u_x(y)$ for each polymer in each microchannel, and we considered two different polymer distributions: the measured distribution $p(y) \propto c(y)$ and a uniform distribution that excluded the polymer center of mass from the space within $R_{\rm g}$ of the wall ($|y| \le W/2-R_{\rm g}$). When using the measured $p(y)$, the agreement between the direct measurement and eq.~\eqref{eq:tayloraris} was good (Fig.~\ref{fig:parity}), with the theory somewhat underpredicting the dispersion compared to the measured value. This discrepancy might be due to the neglect of flow- and confinement-induced anisotropy in the diffusion or inaccuracies in some of the model inputs, but the results were still quite satisfactory. In contrast, using the uniform distribution produced significantly worse agreement (Fig.~S10). This analysis emphasizes the importance of considering the inhomogeneous distribution of the polymers in the microchannel, which depends on both their architecture and concentration, and impacts both the velocity of the polymer solution and the dispersion.

\section{Conclusions}
\label{sec:conclusions}
In this work, we simulated the axial dispersion of linear, comb, and star polymers in dilute solutions flowing through a parallel-plate and an expansion--contraction microchannel. We found that the axial dispersion coefficients increased with the nominal solvent flow rate in both microchannels, as expected, but they depended on both the polymer architecture and whether the polymer concentration was fixed at one value for all polymers or fixed relative to the overlap concentration for each polymer. We then showed that the axial dispersion coefficients could be collapsed as a function of the P\'{e}clet number defined using the polymer self-diffusion coefficients and the average solution velocities measured in the microchannels. 

Our simulations highlight that, in contrast to point-like tracer particles, it can be important to consider confinement effects on polymer diffusion, confinement and flow effects on the polymer concentration distribution in the microchannel, and polymer contributions to the flow field when analyzing and interpreting dispersion measurements for polymer solutions. For example, the polymer axial dispersion coefficients in the expansion--contraction microchannel collapsed with those in the parallel-plate microchannel for the conditions studied, although dispersion coefficients of smaller tracer particles were theoretically expected and confirmed in simulations to be systematically larger in the expansion--contraction microchannel. However, we also found that the dispersion coefficients measured in the parallel-plate microchannel were in reasonable agreement with existing dispersion theory \cite{dispersion_brownian} when we used measured values for the diffusion coefficient, polymer concentration profile, and flow field, suggesting a strategy for analyzing dispersion in polymer solutions. These required inputs for the dispersion theory depend on the confinement of the polymers and are challenging to predict or measure experimentally, so simulations can play an important role here.

We note that in this work there were no significant gradients in the polymer concentration in the flow direction because the simulations employed periodic boundary conditions and were conducted at steady state. Our measured dispersion coefficients are hence most applicable to the dynamics of polymers in steady flows. However, in experiments measuring dispersion coefficients \cite{Alizadeh1980}, a volume of solution is typically injected at a known concentration, then dilutes as the solution disperses. The polymer concentration hence becomes transient and position dependent. It may be possible to use measurements taken at steady concentration to interpret this process, at least approximately, so it may be interesting to simulate dispersion in this context in the future.

\section*{Supplementary Material}
See the supplementary material for additional simulation details, results for monomer solutions, additional velocity profiles for polymer solutions, and theoretical predictions for dispersion coefficients in parallel-plate microchannel using approximate polymer distributions.

\section*{Conflicts of interest}
The authors have no conflicts to disclose.

\section*{Data Availability}
The data that support the findings of this study are available from the authors upon reasonable request.

\begin{acknowledgments}
Acknowledgment is made to the donors of the American Chemical Society Petroleum Research Fund for support of this research (Grant Nos.~65334-DNI7 to AS and 66616-DNI9 to MPH).  This material is based on work supported by the National Science Foundation under Award No.~2143864 (AS and TK). This work used Delta at the National Center for Supercomputing Applications through allocation CHM240073 from the Advanced Cyberinfrastructure Coordination Ecosystem: Services \& Support (ACCESS) program \cite{access}, which is supported by National Science Foundation grants \#2138259, \#2138286, \#2138307, \#2137603, and \#2138296.
\end{acknowledgments}

\bibliography{references}

@inproceedings{access,
	title        = {ACCESS: Advancing Innovation: NSF's Advanced Cyberinfrastructure Coordination Ecosystem: Services \& Support},
	author       = {Boerner, Timothy J. and Deems, Stephen and Furlani, Thomas R. and Knuth, Shelley L. and Towns, John},
	year         = 2023,
	booktitle    = {Practice and Experience in Advanced Research Computing 2023: Computing for the Common Good},
	location     = {Portland, OR, USA},
	publisher    = {Association for Computing Machinery},
	address      = {New York, NY, USA},
	series       = {PEARC '23},
	pages        = {173--176},
	doi          = {10.1145/3569951.3597559},
	isbn         = 9781450399852,
	numpages     = 4
}

@article{ajdari_dispersion_microchannels,
	title        = {Hydrodynamic Dispersion in Shallow Microchannels: the Effect of Cross-Sectional Shape},
	author       = {Ajdari, Armand and Bontoux, Nathalie and Stone, Howard A.},
	year         = 2006,
	journal      = {Analytical Chemistry},
	volume       = 78,
	number       = 2,
	pages        = {387--392},
	doi          = {10.1021/ac0508651}
}

@article{Alexandre_corrugated,
	title        = {Effective description of Taylor dispersion in strongly corrugated channels},
	author       = {Alexandre, Arthur and Gu\'erin, Thomas and Dean, David S.},
	year         = 2025,
	journal      = {Physical Review E},
	publisher    = {American Physical Society},
	volume       = 111,
	pages        = {064124},
	doi          = {10.1103/l1tm-n98s},
	issue        = 6,
	numpages     = 15
}

@article{Alizadeh1980,
	title        = {The theory of the {Taylor} dispersion technique for liquid diffusivity measurements},
	author       = {Alizadeh, A. and Nieto de Castro, C. A. and Wakeham, W. A.},
	year         = 1980,
	journal      = {International Journal of Thermophysics},
	volume       = 1,
	number       = 3,
	pages        = {243--284},
	doi          = {10.1007/BF00517126},
	issn         = {1572-9567}
}

@book{Allen,
	title        = {Computer Simulation of Liquids},
	author       = {Allen, Michael P. and Tildesley, Dominic J.},
	year         = 2017,
	publisher    = {Oxford University Press},
	address      = {New York},
	edition      = 2
}

@article{Androver_laminar,
	title        = {Laminar dispersion at low and high Peclet numbers in a sinusoidal microtube: Point-size versus finite-size particles},
	author       = {Adrover, Alessandra and Venditti, Claudia and Giona, Massimiliano},
	year         = 2019,
	journal      = {Physics of Fluids},
	volume       = 31,
	number       = 6,
	pages        = {062003},
	doi          = {10.1063/1.5096971},
	issn         = {1070-6631}
}

@article{Aris,
	title        = {On the dispersion of a solute in a fluid flowing through a tube},
	author       = {Aris, R.},
	year         = 1956,
	journal      = {Proceedings of the Royal Society of London. A. Mathematical and Physical Sciences},
	volume       = 235,
	number       = 1200,
	pages        = {67--77},
	doi          = {10.1098/rspa.1956.0065},
	issn         = {0080-4630}
}

@misc{azplugins,
	howpublished = {https://github.com/mphowardlab/azplugins}
}

@article{Bhaumik2015251,
	title        = {Taylor--Aris dispersion induced by axial variation in velocity profile in patterned microchannels},
	author       = {Soubhik Kumar Bhaumik and Aadithya Kannan and Sunando DasGupta},
	year         = 2015,
	journal      = {Chemical Engineering Science},
	volume       = 134,
	pages        = {251--259},
	doi          = {10.1016/j.ces.2015.04.052},
	issn         = {0009-2509}
}

@article{Bolster_solute_apertures,
	title        = {Solute dispersion in channels with periodically varying apertures},
	author       = {Bolster, Diogo and Dentz, Marco and Le Borgne, Tanguy},
	year         = 2009,
	journal      = {Physics of Fluids},
	volume       = 21,
	number       = 5,
	pages        = {056601},
	doi          = {10.1063/1.3131982},
	issn         = {1070-6631}
}

@article{Brenner_1962,
	title        = {Effect of finite boundaries on the Stokes resistance of an arbitrary particle},
	author       = {Brenner, Howard},
	year         = 1962,
	journal      = {Journal of Fluid Mechanics},
	volume       = 12,
	number       = 1,
	pages        = {35--48},
	doi          = {10.1017/S0022112062000026}
}

@article{Brenner_1980,
	title        = {Dispersion resulting from flow through spatially periodic porous media},
	author       = {Brenner, H.},
	year         = 1980,
	journal      = {Philosophical Transactions of the Royal Society of London, Series A: Mathematical and Physical Sciences},
	volume       = 297,
	number       = 1430,
	pages        = {81--133},
	doi          = {10.1098/rsta.1980.0205},
	issn         = {0080-4614}
}

@article{Brenner_brownian,
	title        = {The constrained brownian movement of spherical particles in cylindrical pores of comparable radius: Models of the diffusive and convective transport of solute molecules in membranes and porous media},
	author       = {Howard Brenner and Lawrence J Gaydos},
	year         = 1977,
	journal      = {Journal of Colloid and Interface Science},
	volume       = 58,
	number       = 2,
	pages        = {312--356},
	doi          = {10.1016/0021-9797(77)90147-3},
	issn         = {0021-9797}
}

@inbook{Burchard1999SolutionPO,
	title        = {Solution Properties of Branched Macromolecules},
	author       = {Burchard, Walther},
	year         = 1999,
	booktitle    = {Branched Polymers II},
	publisher    = {Springer},
	pages        = {113--194},
	doi          = {10.1007/3-540-49780-3_3}
}

@article{Chang2022TaylorDI,
	title        = {Taylor dispersion in arbitrarily shaped axisymmetric channels},
	author       = {Chang, Ray and Santiago, Juan G.},
	year         = 2023,
	journal      = {Journal of Fluid Mechanics},
	volume       = 976,
	pages        = {A30},
	doi          = {10.1017/jfm.2023.504}
}

@article{Chelakkot_2011,
	title        = {Semiflexible polymer conformation, distribution and migration in microcapillary flows},
	author       = {Chelakkot, Raghunath and Winkler, Roland G and Gompper, Gerhard},
	year         = 2011,
	journal      = {Journal of Physics: Condensed Matter},
	volume       = 23,
	number       = 18,
	pages        = 184117,
	doi          = {10.1088/0953-8984/23/18/184117}
}

@article{chen_dynamics_dna,
	title        = {Conformation and dynamics of single DNA molecules in parallel-plate slit microchannels},
	author       = {Chen, Y.-L. and Graham, M. D. and de Pablo, J. J. and Randall, G. C. and Gupta, M. and Doyle, P. S.},
	year         = 2004,
	journal      = {Physical Review E},
	publisher    = {American Physical Society},
	volume       = 70,
	pages        = {060901(R)},
	doi          = {10.1103/PhysRevE.70.060901},
	issue        = 6,
	numpages     = 4
}

@article{Cottet_2010,
	title        = {Determination of Individual Diffusion Coefficients in Evolving Binary Mixtures by Taylor Dispersion Analysis: Application to the Monitoring of Polymer Reaction},
	author       = {Cottet, Herv{\'e} and Biron, Jean-Philippe and Cipelletti, Luca and Matmour, Rachid and Martin, Michel},
	year         = 2010,
	journal      = {Analytical Chemistry},
	volume       = 82,
	number       = 5,
	pages        = {1793--1802},
	doi          = {10.1021/ac902397x}
}

@article{datta_dispersion_microfluidics,
	title        = {Characterizing dispersion in microfluidic channels},
	author       = {Datta, Subhra and Ghosal, Sandip},
	year         = 2009,
	journal      = {Lab on a Chip},
	publisher    = {The Royal Society of Chemistry},
	volume       = 9,
	pages        = {2537--2550},
	doi          = {10.1039/B822948C},
	issue        = 17
}

@article{Dentz_mixing,
	title        = {Mixing and spreading in stratified flow},
	author       = {Dentz, Marco and Carrera, Jesus},
	year         = 2007,
	journal      = {Physics of Fluids},
	volume       = 19,
	number       = 1,
	pages        = {017107},
	doi          = {10.1063/1.2427089},
	issn         = {1070-6631}
}

@article{Dettmer_diffusion,
	title        = {Anisotropic diffusion of spherical particles in closely confining microchannels},
	author       = {Dettmer, Simon L. and Pagliara, Stefano and Misiunas, Karolis and Keyser, Ulrich F.},
	year         = 2014,
	journal      = {Physical Review E},
	publisher    = {American Physical Society},
	volume       = 89,
	pages        = {062305},
	doi          = {10.1103/PhysRevE.89.062305},
	issue        = 6,
	numpages     = 6
}

@article{diffusion_dna_strychalski,
	title        = {Diffusion of DNA in Nanoslits},
	author       = {Strychalski, Elizabeth A. and Levy, Stephen L. and Craighead, Harold G.},
	year         = 2008,
	journal      = {Macromolecules},
	volume       = 41,
	number       = 20,
	pages        = {7716--7721},
	doi          = {10.1021/ma801313w}
}

@article{dispersion_brownian,
	title        = {Axial dispersion of Brownian colloids in microfluidic channels},
	author       = {Howard, Michael P. and Gautam, Aishwarya and Panagiotopoulos, Athanassios Z. and Nikoubashman, Arash},
	year         = 2016,
	journal      = {Physical Review Fluids},
	publisher    = {American Physical Society},
	volume       = 1,
	pages        = {044203},
	doi          = {10.1103/PhysRevFluids.1.044203},
	issue        = 4,
	numpages     = 20
}

@article{Doi_Harden,
	title        = {Diffusion of macromolecules in narrow capillaries},
	author       = {Harden, J. L. and Doi, M.},
	year         = 1992,
	journal      = {The Journal of Physical Chemistry},
	volume       = 96,
	number       = 10,
	pages        = {4046--4052},
	doi          = {10.1021/j100189a025}
}

@article{eral_anisotropic,
	title        = {Anisotropic and Hindered Diffusion of Colloidal Particles in a Closed Cylinder},
	author       = {Eral, H. B. and Oh, J. M. and van den Ende, D. and Mugele, F. and Duits, M. H. G.},
	year         = 2010,
	journal      = {Langmuir},
	volume       = 26,
	number       = 22,
	pages        = {16722--16729},
	doi          = {10.1021/la102273n}
}

@article{Frankel_Brenner_1989,
	title        = {On the foundations of generalized Taylor dispersion theory},
	author       = {Frankel, I. and Brenner, H.},
	year         = 1989,
	journal      = {Journal of Fluid Mechanics},
	volume       = 204,
	pages        = {97--119},
	doi          = {10.1017/S0022112089001679}
}

@incollection{gompper_2008_mpcd_review,
	title        = {Multi-Particle Collision Dynamics: A Particle-Based Mesoscale Simulation Approach to the Hydrodynamics of Complex Fluids},
	author       = {Gompper, Gerhard and Ihle, Thomas and Kroll, Daniel M. and Winkler, Roland G.},
	year         = 2009,
	booktitle    = {Advanced Computer Simulation Approaches for Soft Matter Sciences III},
	publisher    = {Springer},
	address      = {Berlin, Heidelberg},
	series       = {Advances in Polymer Science},
	volume       = 221,
	pages        = {1--87},
	doi          = {10.1007/978-3-540-87706-6_1},
	editor       = {Holm, Christian and Kremer, Kurt}
}

@article{Hoagland_1985,
	title        = {Taylor-aris dispersion arising from flow in a sinusoidal tube},
	author       = {Hoagland, D. A. and Prud'Homme, R. K.},
	year         = 1985,
	journal      = {AIChE Journal},
	volume       = 31,
	number       = 2,
	pages        = {236--244},
	doi          = {10.1002/aic.690310210}
}

@article{hoomd,
	title        = {HOOMD-blue: A Python package for high-performance molecular dynamics and hard particle Monte Carlo simulations},
	author       = {J. A. Anderson and J. Glaser and S. C. Glotzer},
	year         = 2020,
	journal      = {Computational Materials Science},
	volume       = 173,
	pages        = 109363,
	doi          = {10.1016/j.commatsci.2019.109363}
}

@article{howard_modeling_2019,
	title        = {Modeling hydrodynamic interactions in soft materials with multiparticle collision dynamics},
	author       = {Howard, Michael P and Nikoubashman, Arash and Palmer, Jeremy C},
	year         = 2019,
	journal      = {Current Opinion in Chemical Engineering},
	volume       = 23,
	pages        = {34--43},
	doi          = {10.1016/j.coche.2019.02.007}
}

@article{huang_thermostat_mpcd,
	title        = {Thermostat for nonequilibrium multiparticle-collision-dynamics simulations},
	author       = {Huang, Chien-Cheng and Varghese, Anoop and Gompper, Gerhard and Winkler, Roland G.},
	year         = 2015,
	journal      = {Physical Review E},
	publisher    = {American Physical Society},
	volume       = 91,
	pages        = {013310},
	doi          = {10.1103/PhysRevE.91.013310},
	issue        = 1
}

@article{Ibrahim2013,
	title        = {Size and charge characterization of polymeric drug delivery systems by Taylor dispersion analysis and capillary electrophoresis},
	author       = {Ibrahim, Amal and Meyrueix, R{\'e}mi and Pouliquen, Gauthier and Chan, You Ping and Cottet, Herv{\'e}},
	year         = 2013,
	journal      = {Analytical and Bioanalytical Chemistry},
	volume       = 405,
	number       = 16,
	pages        = {5369--5379},
	doi          = {10.1007/s00216-013-6972-4}
}

@article{ihle_2001_galilean_srd,
	title        = {Stochastic rotation dynamics: A Galilean-invariant mesoscopic model for fluid flow},
	author       = {Ihle, T. and Kroll, D. M.},
	year         = 2001,
	journal      = {Physical Review E},
	volume       = 63,
	pages        = {020201},
	doi          = {10.1103/PhysRevE.63.020201},
	issue        = 2,
	numpages     = 4
}

@article{James_finite,
	title        = {Effective velocity and effective dispersion coefficient for finite-sized particles flowing in a uniform fracture},
	author       = {Scott C. James and Constantinos V. Chrysikopoulos},
	year         = 2003,
	journal      = {Journal of Colloid and Interface Science},
	volume       = 263,
	number       = 1,
	pages        = {288--295},
	doi          = {10.1016/S0021-9797(03)00254-6},
	issn         = {0021-9797}
}

@inbook{Kapral_review,
	title        = {Multiparticle Collision Dynamics: Simulation of Complex Systems on Mesoscales},
	author       = {Kapral, Raymond},
	year         = 2008,
	booktitle    = {Advances in Chemical Physics},
	publisher    = {John Wiley \& Sons, Ltd},
	pages        = {89--146},
	doi          = {10.1002/9780470371572.ch2},
	isbn         = 9780470371572
}

@article{Kg_original,
	title        = {Crossover from Rouse to Reptation Dynamics: A Molecular-Dynamics Simulation},
	author       = {Kremer, Kurt and Grest, Gary S. and Carmesin, I.},
	year         = 1988,
	journal      = {Physical Review Letters},
	publisher    = {American Physical Society},
	volume       = 61,
	pages        = {566--569},
	doi          = {10.1103/PhysRevLett.61.566},
	issue        = 5,
	numpages     = {0}
}

@article{Kitanidis1997,
	title        = {Stokes Flow in a Slowly Varying Two-Dimensional Periodic Pore},
	author       = {Kitanidis, Peter K. and Dykaar, Bruce B.},
	year         = 1997,
	journal      = {Transport in Porous Media},
	volume       = 26,
	number       = 1,
	pages        = {89--98},
	doi          = {10.1023/A:1006575028391}
}

@article{Lamura2001b,
	title        = {Multi-particle collision dynamics: Flow around a circular and a square cylinder},
	author       = {Lamura, A. and Gompper, G. and Ihle, T. and Kroll, D. M.},
	year         = 2001,
	journal      = {Europhysics Letters},
	volume       = 56,
	number       = 3,
	pages        = {319--325},
	doi          = {10.1209/epl/i2001-00522-9}
}

@article{Li_star_branched_dpd,
	title        = {Transport of Star-Branched Polymers in Nanoscale Pipe Channels Simulated with Dissipative Particle Dynamics Simulation},
	author       = {Li, Ziqi and Li, Yajie and Wang, Yongmei and Sun, Zhaoyan and An, Lijia},
	year         = 2010,
	journal      = {Macromolecules},
	volume       = 43,
	number       = 13,
	pages        = {5896--5903},
	doi          = {10.1021/ma100734r}
}

@article{Lin_isolated_sphere,
	title        = {Direct measurements of constrained Brownian motion of an isolated sphere between two walls},
	author       = {Lin, Binhua and Yu, Jonathan and Rice, Stuart A.},
	year         = 2000,
	journal      = {Physical Review E},
	publisher    = {American Physical Society},
	volume       = 62,
	pages        = {3909--3919},
	doi          = {10.1103/PhysRevE.62.3909},
	issue        = 3,
	numpages     = {0}
}

@article{Liu2020RoleOF,
	title        = {Role of Functionality in Cross-Stream Migration, Structures, and Dynamics of Star Polymers in Poiseuille Flow},
	author       = {Liu, Aiqing and Yang, Zhenyue and Liu, Lijun and Chen, Jizhong and An, Lijia},
	year         = 2020,
	journal      = {Macromolecules},
	volume       = 53,
	number       = 22,
	pages        = {9993--10004},
	doi          = {10.1021/acs.macromol.0c00699}
}

@article{Locascio2004,
	title        = {Microfluidic mixing},
	author       = {Locascio, Laurie E.},
	year         = 2004,
	journal      = {Analytical and Bioanalytical Chemistry},
	volume       = 379,
	number       = 3,
	pages        = {325--327},
	doi          = {10.1007/s00216-004-2630-1},
	issn         = {1618-2650}
}

@article{Mai_combs,
	title        = {Conformation of a Comb-like Chain in Solution: Effect of Backbone Rigidity},
	author       = {Mai, Xinghong and Hao, Peng and Liu, Danfeng and Ding, Mingming},
	year         = 2023,
	journal      = {ACS Omega},
	volume       = 8,
	number       = 12,
	pages        = {11177--11183},
	doi          = {10.1021/acsomega.2c08018}
}

@article{malevanets_1999_mesoscopic,
	title        = {Mesoscopic model for solvent dynamics},
	author       = {Malevanets, Anatoly and Kapral, Raymond},
	year         = 1999,
	journal      = {Journal of Chemical Physics},
	volume       = 110,
	number       = 17,
	pages        = {8605--8613},
	doi          = {10.1063/1.478857}
}

@article{malevanets_2000,
	title        = {Dynamics of short polymer chains in solution},
	author       = {A. Malevanets and J. M. Yeomans},
	year         = 2000,
	journal      = {Europhysics Letters},
	volume       = 52,
	number       = 2,
	pages        = 231,
	doi          = {10.1209/epl/i2000-00428-0}
}

@article{mason:physrevlett:1995,
	title        = {Optical Measurements of Frequency-Dependent Linear Viscoelastic Moduli of Complex Fluids},
	author       = {Mason, T. G. and Weitz, D. A.},
	year         = 1995,
	journal      = {Physical Review Letters},
	volume       = 74,
	number       = 7,
	pages        = {1250--1253},
	doi          = {10.1103/PhysRevLett.74.1250}
}

@article{migration_depablo,
	title        = {Cross-Stream Migration of Flexible Molecules in a Nanochannel},
	author       = {Khare, Rajesh and Graham, Michael D. and de Pablo, Juan J.},
	year         = 2006,
	journal      = {Physical Review Letters},
	publisher    = {American Physical Society},
	volume       = 96,
	pages        = 224505,
	doi          = {10.1103/PhysRevLett.96.224505},
	issue        = 22,
	numpages     = 4
}

@article{mussawisade_solvent,
	title        = {Dynamics of polymers in a particle-based mesoscopic solvent},
	author       = {Mussawisade, K. and Ripoll, M. and Winkler, R. G. and Gompper, G.},
	year         = 2005,
	journal      = {The Journal of Chemical Physics},
	volume       = 123,
	number       = 14,
	pages        = 144905,
	doi          = {10.1063/1.2041527},
	issn         = {0021-9606}
}

@article{newtonian_svolent_flow,
	title        = {Mesoscale particle-based simulations of flow in expansion–contraction microchannels at low Reynolds number},
	author       = {Koulaxizis, Tzortzis and De La Torre Garcia, Clara and Petix, C. Levi and Statt, Antonia and Howard, Michael P.},
	year         = 2025,
	journal      = {The Journal of Chemical Physics},
	volume       = 163,
	number       = 19,
	pages        = 194907,
	doi          = {10.1063/5.0299726},
	issn         = {0021-9606}
}

@article{Nikoubashman_branched,
	title        = {Branched Polymers under Shear},
	author       = {Nikoubashman, Arash and Likos, Christos N.},
	year         = 2010,
	journal      = {Macromolecules},
	volume       = 43,
	number       = 3,
	pages        = {1610--1620},
	doi          = {10.1021/ma902212s}
}

@article{Nikoubashman_chain_star_mixtures,
	title        = {Flow Behavior of Chain and Star Polymers and Their Mixtures},
	author       = {Srivastva, Deepika and Nikoubashman, Arash},
	year         = 2018,
	month        = {Jun},
	journal      = {Polymers},
	volume       = 10,
	number       = 6,
	pages        = 599,
	doi          = {10.3390/polym10060599}
}

@article{Nikoubashman_filter,
	title        = {Topology-Sensitive Microfluidic Filter for Polymers of Varying Stiffness},
	author       = {Weiss, Lisa B. and Nikoubashman, Arash and Likos, Christos N.},
	year         = 2017,
	journal      = {ACS Macro Letters},
	volume       = 6,
	number       = 12,
	pages        = {1426--1431},
	doi          = {10.1021/acsmacrolett.7b00768}
}

@article{Ollila_filtration,
	title        = {Biopolymer Filtration in Corrugated Nanochannels},
	author       = {Ollila, Santtu T. T. and Denniston, Colin and Karttunen, Mikko and Ala-Nissila, Tapio},
	year         = 2014,
	journal      = {Physical Review Letters},
	publisher    = {American Physical Society},
	volume       = 112,
	pages        = 118301,
	doi          = {10.1103/PhysRevLett.112.118301},
	issue        = 11,
	numpages     = 5
}

@article{park_dispersion,
	title        = {Dispersion of Flexible Polymer Chains in Confined Geometries},
	author       = {Park, O. Ok},
	year         = 1986,
	journal      = {Korean Journal of Chemical Engineering},
	volume       = 3,
	number       = 2,
	pages        = {153--163},
	doi          = {10.1007/BF02705027},
	issn         = {1975-7220}
}

@article{Ripoll_dynamic,
	title        = {Dynamic regimes of fluids simulated by multiparticle-collision dynamics},
	author       = {Ripoll, M. and Mussawisade, K. and Winkler, R. G. and Gompper, G.},
	year         = 2005,
	journal      = {Physical Review E},
	publisher    = {American Physical Society},
	volume       = 72,
	pages        = {016701},
	doi          = {10.1103/PhysRevE.72.016701},
	issue        = 1,
	numpages     = 14
}

@article{Ripoll_shear,
	title        = {Star Polymers in Shear Flow},
	author       = {Ripoll, Marisol and Winkler, R. and Gompper, Gerhard},
	year         = 2006,
	journal      = {Physical Review Letters},
	volume       = 96,
	pages        = 188302,
	doi          = {10.1103/PhysRevLett.96.188302}
}

@article{Ripoll_star,
	title        = {Hydrodynamic screening of star polymers in shear flow},
	author       = {Ripoll, Marisol and Winkler, R. and Gompper, Gerhard},
	year         = 2007,
	journal      = {The European Physical Journal E},
	volume       = 23,
	pages        = {349--54},
	doi          = {10.1140/epje/i2006-10220-0}
}

@article{Rosecrans_1997,
	title        = {Taylor Dispersion in Curved Channels},
	author       = {Rosencrans, Steve},
	year         = 1997,
	journal      = {SIAM Journal on Applied Mathematics},
	volume       = 57,
	number       = 5,
	pages        = {1216--1241},
	doi          = {10.1137/S003613999426990X}
}

@article{Schaefgen_viscosities,
	title        = {Synthesis of Multichain Polymers and Investigation of their Viscosities1},
	author       = {Schaefgen, John R. and Flory, Paul J.},
	year         = 1948,
	journal      = {Journal of the American Chemical Society},
	volume       = 70,
	number       = 8,
	pages        = {2709--2718},
	doi          = {10.1021/ja01188a026}
}

@article{Silebi_submicron,
	title        = {Axial dispersion of submicron particles in capillary hydrodynamic fractionation},
	author       = {Silebi, Cesar A. and Dosramos, Jose G.},
	year         = 1989,
	journal      = {AIChE Journal},
	volume       = 35,
	number       = 8,
	pages        = {1351--1364},
	doi          = {10.1002/aic.690350814}
}

@article{squires_fluid_physics,
	title        = {Microfluidics: Fluid physics at the nanoliter scale},
	author       = {Squires, Todd M. and Quake, Stephen R.},
	year         = 2005,
	journal      = {Reviews of Modern Physics},
	publisher    = {American Physical Society},
	volume       = 77,
	pages        = {977--1026},
	doi          = {10.1103/RevModPhys.77.977},
	issue        = 3,
	numpages     = {0}
}

@article{stone_microfluidics,
	title        = {Engineering Flows in Small Devices: Microfluidics Toward a Lab-on-a-Chip},
	author       = {Stone, H.A. and Stroock, A.D. and Ajdari, A.},
	year         = 2004,
	journal      = {Annual Review of Fluid Mechanics},
	publisher    = {Annual Reviews},
	volume       = 36,
	number       = {Volume 36, 2004},
	pages        = {381--411},
	doi          = {10.1146/annurev.fluid.36.050802.122124},
	issn         = {1545-4479},
	type         = {Journal Article}
}

@article{Taylor,
	title        = {Dispersion of soluble matter in solvent flowing slowly through a tube},
	author       = {Taylor, Geoffrey Ingram},
	year         = 1953,
	journal      = {Proceedings of the Royal Society of London. A. Mathematical and Physical Sciences},
	volume       = 219,
	number       = 1137,
	pages        = {186--203},
	doi          = {10.1098/rspa.1953.0139},
	issn         = {0080-4630}
}

@article{Taylor_1954,
	title        = {Conditions under which dispersion of a solute in a stream of solvent can be used to measure molecular diffusion},
	author       = {Taylor, Geoffrey Ingram},
	year         = 1954,
	journal      = {Proceedings of the Royal Society of London. A. Mathematical and Physical Sciences},
	volume       = 225,
	number       = 1163,
	pages        = {473--477},
	doi          = {10.1098/rspa.1954.0216},
	issn         = {0080-4630}
}

@article{teodoro2025taylor,
	title        = {Taylor dispersion analysis in a viscoelastic fluid through arbitrarily shaped axisymmetric channels},
	author       = {Teodoro, Carlos and Bautista, Oscar and M\'{e}ndez, Federico},
	year         = 2025,
	journal      = {Journal of Fluid Mechanics},
	volume       = 1025,
	pages        = {A9},
	doi          = {10.1017/jfm.2025.10928}
}

@article{Vartuli_capillary,
	title        = {Taylor dispersion in a polymer solution flowing in a capillary tube},
	author       = {Vartuli, M. and Hulin, J. P. and Daccord, G.},
	year         = 1995,
	journal      = {AIChE Journal},
	volume       = 41,
	number       = 7,
	pages        = {1622--1628},
	doi          = {10.1002/aic.690410703}
}

@article{Virtanen2020,
	title        = {{SciPy} 1.0: fundamental algorithms for scientific computing in {Python}},
	author       = {Virtanen, Pauli and Gommers, Ralf and Oliphant, Travis E. and Haberland, Matt and Reddy, Tyler and Cournapeau, David and Burovski, Evgeni and Peterson, Pearu and Weckesser, Warren and Bright, Jonathan and van der Walt, St{\'e}fan J. and Brett, Matthew and Wilson, Joshua and Millman, K. Jarrod and Mayorov, Nikolay and Nelson, Andrew R. J. and Jones, Eric and Kern, Robert and Larson, Eric and Carey, C. J. and Polat, {\.I}lhan and Feng, Yu and Moore, Eric W. and VanderPlas, Jake and Laxalde, Denis and Perktold, Josef and Cimrman, Robert and Henriksen, Ian and Quintero, E. A. and Harris, Charles R. and Archibald, Anne M. and Ribeiro, Ant{\^o}nio H. and Pedregosa, Fabian and van Mulbregt, Paul and Vijaykumar, Aditya and Bardelli, Alessandro Pietro and Rothberg, Alex and Hilboll, Andreas and Kloeckner, Andreas and Scopatz, Anthony and Lee, Antony and Rokem, Ariel and Woods, C. Nathan and Fulton, Chad and Masson, Charles and H{\"a}ggstr{\"o}m, Christian and Fitzgerald, Clark and Nicholson, David A. and Hagen, David R. and Pasechnik, Dmitrii V. and Olivetti, Emanuele and Martin, Eric and Wieser, Eric and Silva, Fabrice and Lenders, Felix and Wilhelm, Florian and Young, G. and Price, Gavin A. and Ingold, Gert-Ludwig and Allen, Gregory E. and Lee, Gregory R. and Audren, Herv{\'e} and Probst, Irvin and Dietrich, J{\"o}rg P. and Silterra, Jacob and Webber, James T. and Slavi{\v{c}}, Janko and Nothman, Joel and Buchner, Johannes and Kulick, Johannes and Sch{\"o}nberger, Johannes L. and de Miranda Cardoso, Jos{\'e} Vin{\'i}cius and Reimer, Joscha and Harrington, Joseph and Rodr{\'i}guez, Juan Luis Cano and Nunez-Iglesias, Juan and Kuczynski, Justin and Tritz, Kevin and Thoma, Martin and Newville, Matthew and K{\"u}mmerer, Matthias and Bolingbroke, Maximilian and Tartre, Michael and Pak, Mikhail and Smith, Nathaniel J. and Nowaczyk, Nikolai and Shebanov, Nikolay and Pavlyk, Oleksandr and Brodtkorb, Per A. and Lee, Perry and McGibbon, Robert T. and Feldbauer, Roman and Lewis, Sam and Tygier, Sam and Sievert, Scott and Vigna, Sebastiano and Peterson, Stefan and More, Surhud and Pudlik, Tadeusz and Oshima, Takuya and Pingel, Thomas J. and Robitaille, Thomas P. and Spura, Thomas and Jones, Thouis R. and Cera, Tim and Leslie, Tim and Zito, Tiziano and Krauss, Tom and Upadhyay, Utkarsh and Halchenko, Yaroslav O. and V{\'a}zquez-Baeza, Yoshiki and {SciPy 1.0 Contributors}},
	year         = 2020,
	journal      = {Nature Methods},
	volume       = 17,
	number       = 3,
	pages        = {261--272},
	doi          = {10.1038/s41592-019-0686-2}
}

@article{wang2023behaviors,
	title        = {Behaviors of a Polymer Chain in Channels: From Zimm to Rouse Dynamics},
	author       = {Wang, Zhenhua and Wang, Zhen-Gang and Shi, An-Chang and Lu, Yuyuan and An, Lijia},
	year         = 2023,
	journal      = {Macromolecules},
	publisher    = {American Chemical Society},
	volume       = 56,
	number       = 6,
	pages        = {2447--2453},
	doi          = {10.1021/acs.macromol.3c00013}
}

@article{Weeks_wca_1971,
	title        = {Role of Repulsive Forces in Determining the Equilibrium Structure of Simple Liquids},
	author       = {Weeks, John D. and Chandler, David and Andersen, Hans C.},
	year         = 1971,
	journal      = {The Journal of Chemical Physics},
	volume       = 54,
	number       = 12,
	pages        = {5237--5247},
	doi          = {10.1063/1.1674820},
	issn         = {0021-9606}
}

@article{weiss2019hydrodynamics,
	title        = {Hydrodynamics and Filtering of Knotted Ring Polymers in Nanochannels},
	author       = {Weiss, Lisa B. and Marenda, Mattia and Micheletti, Cristian and Likos, Christos N.},
	year         = 2019,
	journal      = {Macromolecules},
	volume       = 52,
	number       = 11,
	pages        = {4111--4119},
	doi          = {10.1021/acs.macromol.9b00516}
}

@article{weiss2019spatial,
	title        = {Spatial Demixing of Ring and Chain Polymers in Pressure-Driven Flow},
	author       = {Weiss, Lisa B. and Likos, Christos N. and Nikoubashman, Arash},
	year         = 2019,
	journal      = {Macromolecules},
	volume       = 52,
	number       = 20,
	pages        = {7858--7869},
	doi          = {10.1021/acs.macromol.9b01629}
}

@article{Whitmer_2010,
	title        = {Fluid--solid boundary conditions for multiparticle collision dynamics},
	author       = {Whitmer, Jonathan K and Luijten, Erik},
	year         = 2010,
	journal      = {Journal of Physics: Condensed Matter},
	volume       = 22,
	number       = 10,
	pages        = 104106,
	doi          = {10.1088/0953-8984/22/10/104106}
}

@article{Yang2020,
	title        = {Role of Hydrodynamic Interactions in the Deformation of Star Polymers in Poiseuille Flow},
	author       = {Yang, Zhen-Yue and Tian, Xiao-Fei and Liu, Li-Jun and Chen, Ji-Zhong},
	year         = 2020,
	journal      = {Chinese Journal of Polymer Science},
	volume       = 38,
	number       = 4,
	pages        = {363--370},
	doi          = {10.1007/s10118-020-2346-5}
}

@article{zimm_branches,
	title        = {The Dimensions of Chain Molecules Containing Branches and Rings},
	author       = {Zimm, Bruno H. and Stockmayer, Walter H.},
	year         = 1949,
	journal      = {The Journal of Chemical Physics},
	volume       = 17,
	number       = 12,
	pages        = {1301--1314},
	doi          = {10.1063/1.1747157},
	issn         = {0021-9606}
}

\end{document}


\title{Supplementary material for ``Axial dispersion in dilute solutions of linear and branched polymers in parallel-plate and expansion--contraction microchannels``}

\author{C. Levi Petix}
\thanks{These authors contributed equally.}
\affiliation{Department of Chemical Engineering, Auburn University, Auburn, AL 36849, USA}

\author{Tzortzis Koulaxizis}
\thanks{These authors contributed equally.}
\affiliation{Department of Chemical and Biomolecular Engineering, University of Illinois, Urbana--Champaign, Illinois 61801, USA}

\author{Griffin D. Overton}
\affiliation{Department of Chemical Engineering, Auburn University, Auburn, AL 36849, USA}

\author{Antonia Statt}
\email{statt@illinois.edu}
\affiliation{Department of Chemical and Biomolecular Engineering, University of Illinois, Urbana--Champaign, Illinois 61801, USA}
\affiliation{Department of Materials Science and Engineering, The Grainger College of Engineering,  University of Illinois, Urbana--Champaign, Illinois 61801, USA}

\author{Michael P. Howard}
\email{mphoward@auburn.edu}
\affiliation{Department of Chemical Engineering, Auburn University, Auburn, AL 36849, USA}

\maketitle

\begin{figure}
    \centering
    \includegraphics{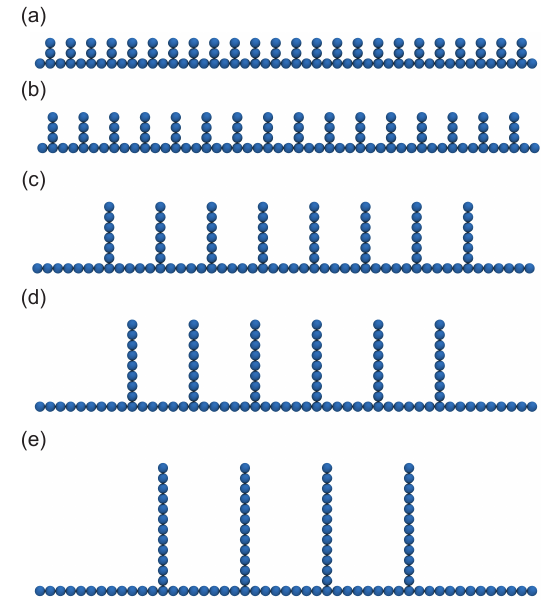}
    \caption{Comb polymers with a total of 97 monomers and number of arms $f=$ (a) 24, (b) 16, (c) 8, (d) 6, and (e) 4.}
\end{figure}

\begin{table}
\setlength{\tabcolsep}{10pt}
\caption{Side length $L$ of cubic simulation box and number of polymers $N$ used to measure the bulk self-diffusion coefficients for each polymer at fixed concentration relative to overlap.}
\begin{tabular}{cccc}
polymer & $f$ & $L$ ($\ell$) & $N$ \\
\hline
linear &  & 100 & 305 \\
star   & 4  & 77  & 308 \\
star   & 6  & 64  & 308 \\
star   & 8  & 56  & 310 \\
comb   & 4  & 75  & 302 \\
comb   & 6  & 74  & 305 \\
comb   & 8  & 74  & 301 \\
comb   & 16 & 77  & 303 \\
comb   & 24 & 76  & 300
\end{tabular}
\end{table}

\begin{table}
\setlength{\tabcolsep}{10pt}
\caption{Number of polymers $N$ in the parallel-plate and expansion--contraction microchannels at fixed concentration relative to overlap. All simulations at fixed concentration had $N = 82$ polymers.}
\begin{tabular}{cc}
polymer & $N$ \\
\hline
linear      & 82  \\
4-arm comb  & 193 \\
8-arm comb  & 200 \\
4-arm star  & 182 \\
8-arm star  & 477
\end{tabular}
\end{table}

\begin{table}
\setlength{\tabcolsep}{10pt}
\caption{Body force $F_x$ (in $10^{-5}\,\varepsilon/\ell$) applied in the $x$ direction for each nominal volumetric flow rate $Q$ (in $\ell^3/\tau$) in the parallel-plate and expansion--contraction microchannels.}
\begin{tabular}{ccc}
$Q$ & $F_x$, parallel plate & $F_x$, expansion--contraction \\
\hline
5  & 0.804 & 1.49 \\
10 & 1.61 & 2.98 \\
15 & 2.41 & 4.47 \\
20 & 3.21 & 5.96 \\
25 & 4.02 & 7.44
\end{tabular}
\end{table}

\begin{figure}
  \centering
  \rotatebox{90}{
    \begin{minipage}{\textheight}
      \centering
      \includegraphics[width=\linewidth]{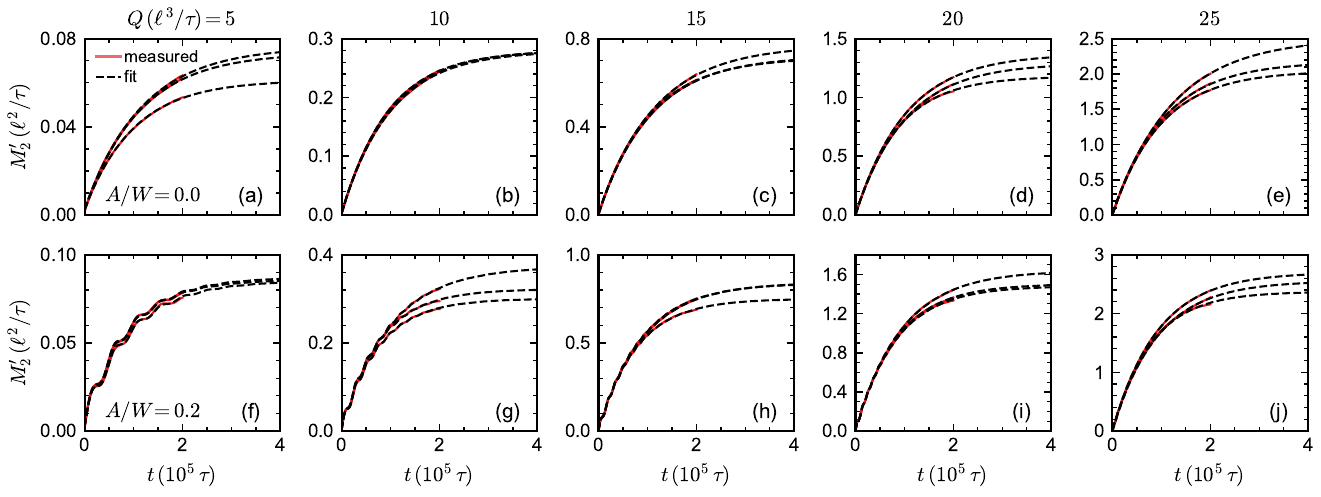}
      \caption{Representative measured values and fits used to extract the axial dispersion coefficient $K$ from the time-dependent dispersion data. Measured values of $M_2'$ and corresponding fits are shown for all 3 independent simulations in the (a)--(e) parallel-plate microchannel ($A/W = 0.0$) and (f)--(j) expansion--contraction microchannel ($A/W = 0.2$) at nominal volumetric flow rates $Q=5$, 10, 15, 20, and $25\,\ell^3/\tau$. The fits in (a)--(e) have an exponential form [eq.~(6)], while those in (f)--(j) have a damped oscillatory form [eq.~(7)] except for panel (j), which uses the exponential form.}
    \end{minipage}
  }
\end{figure}

\begin{figure}
    \centering
    \includegraphics{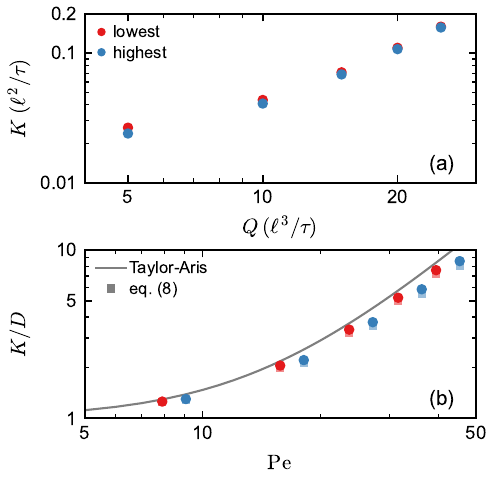}
    \caption{Dispersion of nonbonded monomers in the parallel-plate microchannel at the lowest and highest monomer concentrations considered. (a) Axial dispersion coefficient $K$ as a function of nominal volumetric flow rate $Q$. (b) Dimensionless axial dispersion coefficient $K/D$ as a function of P\'{e}clet number Pe. The solid line shows the classical Taylor--Aris prediction, while the squares show the prediction of the extended theory given by eq.~(8) using the measured diffusion coefficient in the microchannel (Fig.~\ref{fig:monomer_diffusion}), measured concentration profile (Fig.~\ref{fig:monomer_concentration_profiles}), and measured velocity (Fig.~\ref{fig:velocity_monomers}).}
\end{figure}

\begin{figure}
    \centering
    \includegraphics{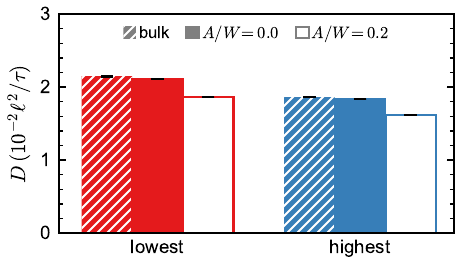}
    \caption{Self-diffusion coefficient $D$ of nonbonded monomers at the lowest and highest monomer concentrations considered, measured in bulk solution, the parallel-plate microchannel ($A/W=0.0$), and the expansion--contraction microchannel ($A/W=0.2$).}
    \label{fig:monomer_diffusion}
\end{figure}

\begin{figure}
    \centering
    \includegraphics{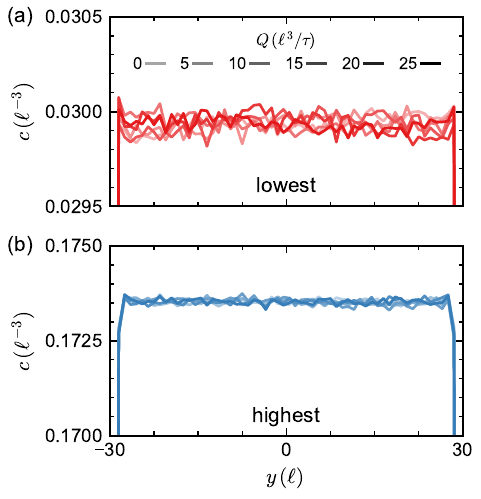}
    \caption{Concentration profile $c(y)$ of nonbonded monomers across the parallel-plate microchannel at the (a) lowest and (b) highest monomer concentrations considered at nominal volumetric flow rates $Q=0$, 5, 10, 15, 20, and $25\,\ell^3/\tau$.}
    \label{fig:monomer_concentration_profiles}
\end{figure}

\begin{figure}
    \centering
    \includegraphics{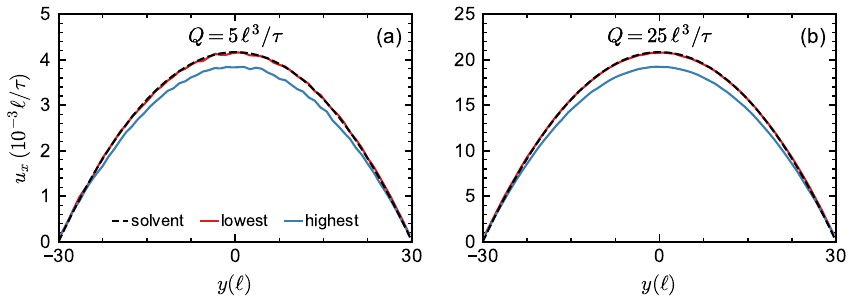}
    \caption{Axial velocity profile $u_x(y)$ in the parallel-plate microchannel for nonbonded monomers at the lowest and highest monomer concentrations considered for $Q=$ (a) $5\,\ell^3/\tau$ and (b) $25 \, \ell^3/\tau$. The expected behavior for the pure solvent is shown as a dashed black line for reference.}
    \label{fig:velocity_monomers}
\end{figure}

\begin{figure}
    \centering
    \includegraphics{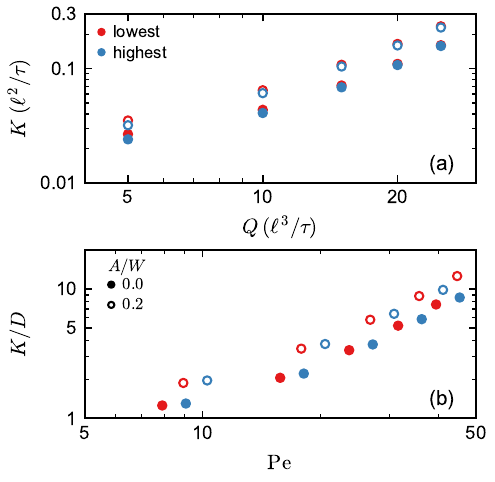}
    \caption{Dispersion of nonbonded monomers in the expansion--contraction microchannel ($A/W = 0.2$) compared to that in the parallel-plate microchannel ($A/W = 0.0$) at the lowest and highest monomer concentrations considered. (a) Axial dispersion coefficient $K$ as a function of nominal volumetric flow rate $Q$. (b) Dimensionless axial dispersion coefficient $K/D$ as a function of P\'{e}clet number ${\rm Pe}$.}
\end{figure}

\begin{figure}
    \centering
    \includegraphics{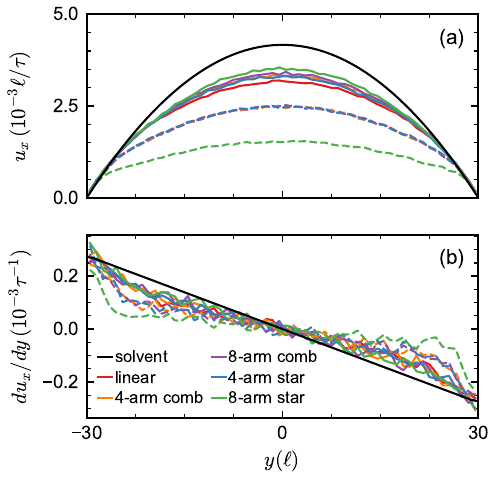}
    \caption{Same as Fig.~6 but at $Q=5\,\ell^3/\tau$.}
\end{figure}

\begin{figure}
    \centering
    \includegraphics{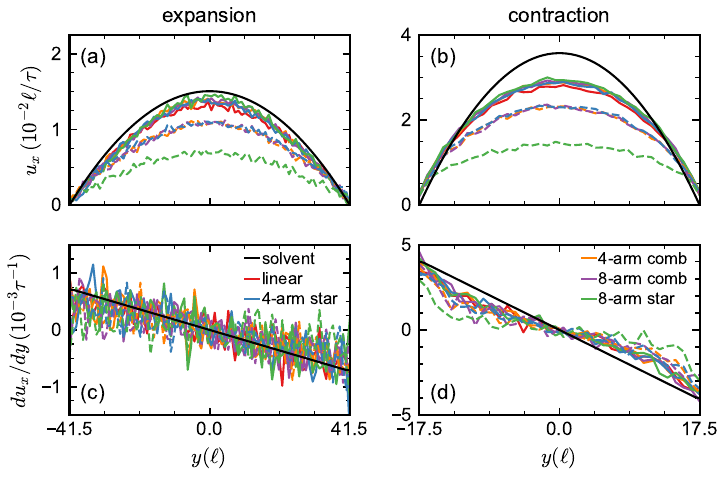}
    \caption{Same as Fig.~6 but for the expansion--contraction microchannel at the (a), (c) expansion and (b), (d) contraction points.}
\end{figure}

\begin{figure}
    \centering
    \includegraphics{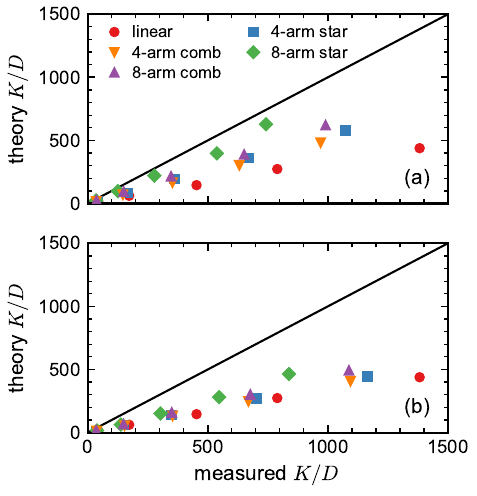}
    \caption{Same as Fig.~8 but assuming a uniform distribution of the polymer for $|y| \le W/2-R_{\rm g}$.}
\end{figure}